\documentclass[preprint,12pt,
aps,pra,
longbibliography,
floatfix,
groupedaddress,superscriptaddress
]{revtex4-1}

\usepackage{graphicx}
\usepackage{dcolumn}
\usepackage{bm}
\usepackage{color}
\usepackage{natbib}
\usepackage[margin=1in]{geometry}
\usepackage{color}
\usepackage[dvipsnames]{xcolor}

\begin{document}

\title{
Optical Detection and Storage of Entanglement in Plasmonically Coupled Quantum Dot Qubits}

\author{M.~Otten}
\affiliation{Center for Nanoscale Materials, Argonne National Laboratory, Argonne, IL 60439, USA.}
\author{S.~K.~Gray}
\affiliation{Center for Nanoscale Materials, Argonne National Laboratory, Argonne, IL 60439, USA.}
\author{G.~V.~Kolmakov}
\affiliation{New York City College of Technology, the City University of New York, Brooklyn, NY 11201, USA.
}


\begin{abstract}
Recent proposals and advances in quantum simulations, quantum cryptography and quantum communications substantially rely on quantum entanglement formation.
Contrary to the conventional wisdom that dissipation destroys quantum coherence, coupling with a dissipative environment can also generate entanglement.
We consider a system composed of two quantum dot qubits coupled with a common,  damped surface plasmon mode; each quantum dot is also coupled to a separate photonic cavity mode. Cavity quantum electrodynamics calculations show that upon optical excitation by
a femtosecond laser pulse,  entanglement of the quantum dot excitons occurs, and the time evolution of the $g^{(2)}$ pair correlation function of the cavity photons is an indicator of the 
entanglement. We also show that the degree of entanglement is conserved during the time evolution of the system. Furthermore, if coupling of the photonic cavity and quantum dot modes is large enough,  the quantum dot entanglement can be transferred to the cavity modes to increase the overall entanglement lifetime.
This latter phenomenon can be viewed as a signature of entangled, long-lived quantum dot exciton-polariton formation.
The preservation of total entanglement in the strong coupling limit of the cavity/quantum dot interactions suggests a novel means of  entanglement storage and manipulation in high-quality optical cavities.
\end{abstract}


\maketitle

\section{INTRODUCTION}

In the past decade, rapid developments in 
quantum cryptography, quantum communications,  and quantum simulations (Refs.\ \cite{Gisin:07,Thearle:18,Buluta:08} 
and references therein) have been made. Interesting examples include
the reported low-Earth-orbit satellite-to-ground quantum state transmission \cite{Liao:17} 
that paves the route to a secure ``quantum internet,''
a hundred-kilometer-long optical line for quantum key distribution \cite{Schmitt:07,Varnava:16}, and a
publicly available 20-qubit universal quantum computer \cite{ibmq:17}. 
These  and other advances  are stimulating further research into  systems that can provide physical realizations of quantum 
entangled states \cite{Reilly:15,Pan:12}. Previously proposed techniques include coupled quantum rings \cite{OConnor:01},
quasi-phase-matching ring crystals for entangled photon generation \cite{Hua:18}, and  subradiant 
Dicke states of trapped interacting atoms \cite{Dicke:54,Ficek:02}. 
Recently, a solid-state realization of trapped atoms   -- coupled quantum dot (QD) qubits, or ``artificial molecules,'' 
in   a dissipative environment  --  has been proposed to provide  entanglement among  electron-hole excitations in QDs (i.e., excitons)
in two- and multi-dot systems at liquid helium temperatures \cite{Martin-Cano:11,Hou:14,Otten:15,Otten:16,Thorgrimsson:17}.
These predictions, still to be validated experimentally,  potentially open a new route to  
the design of robust solid-state emitters of entangled photons 
of relevance to quantum information science and sensing \cite{Hu:00,Ladd:10,Jayakumar:14}. 

Some time ago  Burkard, Loss and DiVincenzo \cite{Burkard:99} proposed coupled quantum dots as a platform for the design of quantum gates, to be used in prospective quantum computers.
It is worth noting that the coupled quantum dot system in high-quality cavity can be mapped into cavity quantum electrodynamics of superconducting electrical circuits \cite{Blais:04}, which is one of the promising architectures for  quantum simulations. Very recently, a programmable two-qubit quantum processor has been realized based on two quantum dots in silicon \cite{Watson:18}.

It has been found that the second-order correlation function of photons emitted by coupled QD qubits
can  be utilized as a ``witness'' of quantum entanglement formation; specifically, 
antibunching of photons emitted by the QDs has been numerically predicted \cite{Dumitrescu:17}. 
Various QD coupling methods have been proposed including   
sharing the photon field in an optical microcavity \cite{Gies:15}
or the interactions with  auxiliary plasmonic nanoantennas \cite{Martin-Cano:11,Hou:14,Otten:15,Otten:16}.
However, direct observations of the entanglement of electronic degrees of freedom in QDs remain challenging.

In this work, we suggest a new system for achieving, detecting and, further, manipulating entanglement of plasmonically coupled QDs in optical microcavities. Specifically,
through numerical simulations  and theoretical analysis, we demonstrate that, under suitable circumstances, there can be a 
 one-to-one correspondence between the entanglement of QD states  and the
correlation properties of cavity photons emitted  by the QDs.
Further, we show that  the time dependence of both the QD entanglement and the photon correlation functions 
is drastically changed from  photon correlation suppression or antibunching to strong oscillations 
during the transition from the weak to strong QD-cavity photon coupling regimes. 
The latter oscillations are the signature of entangled 
exciton-polariton states, in which the quantum correlations are shared between the QD excitons and cavity photons.
In our work, we consider a quantum system driven by a strong femtosecond laser pump. 
Our results will enable the identification of entanglement in coupled QD systems via 
cavity photon correlation measurements and also suggest a means of
storing such entanglement in the cavity modes.

It is important to note that there is no simple rule -- bunching vs antibunching -- to be associated with the photon pair correlation function and entanglement or strong coupling and the
results will depend on the specific systems and measurements carried out. 
Here we show, in the system proposed and in the limit of strong coupling between QDs and photonic cavities, that QD entanglement manifests itself as photon {\it bunching} or sharp peaks in the same-time cross-correlation function for the photons.
Thus our
results can be contrasted
 with the steady-state antibunching correlation noted by Dumitrescu and Lawrie \cite{Dumitrescu:17} that would occur in a quantum dot/plasmon system without coupling to photonic cavities.  The underlying mechanism that leads to the bunching/anti-bunching behavior in our case is also fundamentally different from that which leads to the bunching behavior predicted for coupled plasmonic systems due to their bosonic character by Masiello and co-workers \cite{Thakkar:14} and the ultra-strong coupling bunching behavior noted by Savasta and co-workers \cite{Garziano:17}.  

\section{Numerical model}

\subsection{Cavity Quantum Electrodynamics}

\begin{figure}[t]
\begin{center}
\includegraphics[width=140.mm]{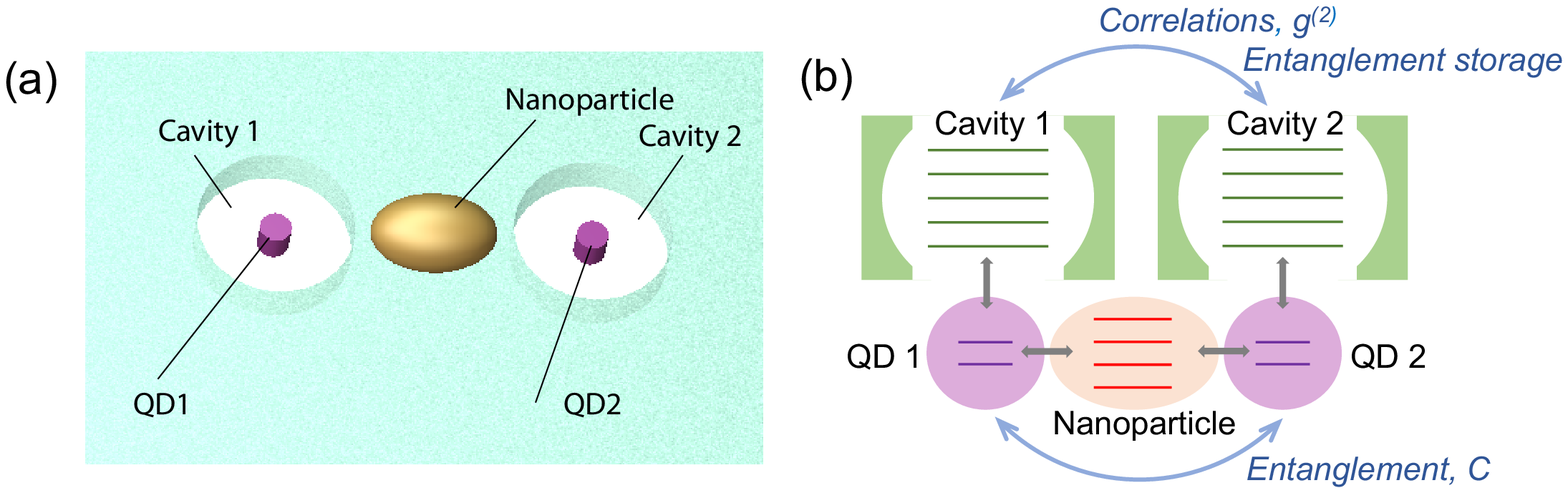}
\end{center} 
\caption{(Color online) Schematic of the system. {(a)} Two quantum dot (QD) qubits are embedded into optical cavities and coupled 
with plasmonic modes in a neighboring metal nanoparticle. The chosen 
setup with two individual cavities, each of them encloses a single QD, enables one to separately
 tune the QD-photon and QD-plasmon coupling strengths.
{(b)}  Graphical representation of our model: 
the two-level QDs are coupled with  plasmonic modes and  photon cavity modes; gray arrows shows the respective coupling.
In our calculations, the QDs and plasmons are excited by a  laser pulse with full width at half maximum of 20 fs. We show that QD qubit entanglement, defined as Wootters' concurrence $C$,  can be both detected via the $g^{(2)}$ pair correlation function of the cavity photon and stored in high-quality optical cavities.} 
\label{fig:schematic}
\end{figure}

A schematic of our system is shown in Fig.\ \ref{fig:schematic}.
The system  is composed of two QDs  embedded into optical cavities. 
In addition, the QD electronic degrees of freedom  are coupled with the surface plasmon modes in a neighboring metal particle or nanostructure, 
which provides an efficient dissipative environment.
The chosen scheme enables one to controllably and independently tune the QD coupling strengths, compared 
to the setup where QDs share the same optical cavity. 
The temporal evolution of the whole system is described by the cavity quantum electrodynamics (CQED)
equation for the time-dependent density operator 
$\hat{\rho}(t)$ \cite{Shan:13,Otten:16}, 
\begin{equation}
\frac{\partial \hat{\rho}}{\partial t} = - \frac{i}{\hbar} [ \hat{H},  \hat{\rho} ] - \frac{i}{\hbar} [ \hat{H}_d,  \hat{\rho} ] 
+ L(\hat{\rho}), \label{eq:master}
\end{equation} 
where $\hat{H} = \hat{H}_0 + \hat{H}_{int}$ is the system Hamiltonian that includes the free Hamiltonians of two ($i=1,2$) two-level QDs, surface plasmon and cavity photon modes 
\begin{equation}
\hat{H}_0=\sum_{i}\hbar \omega_i \hat{\sigma}^\dag_i \hat{\sigma}_i + \hbar \omega_s \hat{b}^\dag \hat{b} + \sum_{i}\hbar \omega_i \hat{c}^\dag_i \hat{c}_i,
\end{equation}
their interactions 
\begin{equation}
\hat{H}_{int} = -\sum_i \hbar g_s^i (\hat{\sigma}^\dag_i \hat{b} + \hat{\sigma}_i \hat{b}^\dag) 
-\sum_i \hbar g (\hat{\sigma}^\dag_i \hat{c}_i + \hat{\sigma}_i \hat{c}^\dag_i),
\end{equation}
and coupling 
\begin{equation}
\hat{H}_d = -E(t) \left[ \sum_i d_i (\hat{\sigma}_i +  \hat{\sigma}^\dag_i )
+ d_s (\hat{b} +  \hat{b}^\dag )\right]
\end{equation}
with the external driving electromagnetic field $E(t)$ 
considered in the semiclassical dipole limit (with $\hat{\sigma}_i$, $\hat{b}$ and $\hat{c}_i$
to be the respective annihilation operators for QDs, plasmons and cavity photon excitations);
$d_i$ and $d_s$ are the transition dipole moments of the QDs and plasmons, respectively. 
We emphasize that the annihilation operators in Eq.\ (\ref{eq:master}) act in the coordinate space; thus, whereas the total electron excitation and photon wave functions obey the conventional symmetry dictated by their statistics, their coordinate wave functions can be both symmetric and antisymmetric.
The Lindblad superoperator  $L(\hat{\rho})$ accounts for  the QD and cavity photon 
population relaxation and dephasing and plasmon dissipation \cite{Shan:13,Otten:15}. 

Eq.~(\ref{eq:master})  was numerically solved in the rotating-phase approximation with the 
recently developed  ``Open quantum systems in C'' 
(QuaC) simulation package \cite{QuaC:17} based on sparse matrix-vector multiplication algorithms 
along with the 4th-order Runge-Kutta numerical scheme. 
We found that the results converged for the number of photon levels $N_{ph}=4$ and the number of plasmon levels 
$N_{pl}=24$.
The QD entanglement is captured via Wootters' concurrence $C(t)$ calculated from the QD reduced 
density matrix \cite{Wootters:98}.

We characterize light in the photonic
cavities with the normalized  pair correlation function for 
photons arriving at the same time \cite{Yamamoto:99,Scully:97},
which in our case simplifies to
\begin{equation}
 g_{ij}^{(2)}(t) \equiv  g_{ij}^{(2)}(t,\tau = 0) = 
 \frac{{\rm Tr} ( \hat{c}_i^\dag  \hat{c}_j^\dag  \hat{c}_j \hat{c}_i \hat{\rho} (t) )}{n_i(t) n_j(t)}
\end{equation} 
\noindent where $i,j$ = 1,2 denote the photonic cavity modes, $\hat{c}_i^\dag$, $\hat{c}_i$ are
the corresponding creation/annihilation operators, and  
$n_i(t) = {\rm Tr}(\hat{c}_i^\dag \hat{c}_i \hat{\rho}(t))$ is the time-dependent population of the $i$th photon mode.

\subsection{Relevant Parameters}

In what follows, the QDs are illuminated with a pulsed electric 
field $E(t)=E_0(t)\cos(\omega t)$, where $E_0(t)$ 
is a Gaussian  envelope function with the maximum electric field of 
$2.5\times 10^6$ V/m and the 
full width at half maximum (fwhm) of 20 fs
(a fluence of 26.4 nJ/cm$^2$, see Appendix A).
In our simulations we consider low cavity photon mode occupations so that we can disregard  
renormalization of the photon energies due to nonlinearity \cite{Verger:06}.

In our simulations, we set the dephasing rate to 8.6~$\mu$eV, corresponding to the temperature 0.1~K. (We note that this temperature is about an order of magnitude higher than that used for superconducting qubits \cite{ibmq:17}.)
In our simulations we consider low cavity photon mode occupations so that we can disregard  
renormalization of the photon energies due to nonlinearity \cite{Verger:06}.
Eq.~(\ref{eq:master})  was numerically solved in the rotating-phase approximation with the 
recently developed  ``Open quantum systems in C'' 
(QuaC) simulation package \cite{QuaC:17} based on sparse matrix-vector multiplication algorithms 
along with the 4th-order Runge-Kutta numerical scheme. 
We found that the results converged for the number of photon levels $N_{ph}=4$ and the number of plasmon levels 
$N_{pl}=24$.

The photonic cavity environment presents a powerful means for controlling light matter interactions in solid-state systems \cite{Galfsky:17}.
Variations in the cavity geometry, the cavity-QD energy detuning, and
the QD position relative to the maximum of the light electric field in the cavity and to the plasmonic structure
provide an experimental opportunity to alter the QD-cavity photon and QD-surface plasmon interactions strengths 
in wide limits \cite{Senellart:08,Pelton:15,Chikkaraddy:16}.
Furthermore,
the use of anisotropic metamaterials as environments for the QDs could provide  an additional important tool to manipulate light matter interactions since, as demonstrated by Menon and co-workers,
optical topological transitions in these materials significantly modify photon emission rates \cite{Krishnamoorthy:12}.
The respective QD-photon dimensionless coupling strength $\xi= 4g/(\gamma_{QD}+\gamma_C)$ varies
from $\xi \ll 1$ (weak coupling regime) to $\sim 2$ (strong coupling) \cite{Savona:95,Creatore:08,Hennessy:07,Senellart:08}
(with $\gamma_{QD(C)}$ to be the  population relaxation rates for QD excitons (cavity photons)). 
The latter value is comparable with $\xi \approx 5$ for single atoms in an optical microcavity \cite{Boca:04}. By making use
of the Purcell effect,
$\xi$ can be  further enhanced by an order of magnitude by using  high-finesse optical cavities \cite{Purcell:46,Senellart:08,Pelton:15} with quality 
$Q$-factor $\sim 10^5$.
Thanks to the interaction of QDs with overdamped surface plasmons in the neighboring particle,
the QD relaxation dominates with the  respective decay rate \cite{Shan:13} $\gamma_{QD} = 4 g_s^{2}/\gamma_s > \gamma_C$  
(see also Appendix A). Recently, the utility of the strong qubit-photon coupling regimes has been demonstrated for CQED with flux qubits by Armata {\it et al.} in Ref.\ \cite{Armata:17} where, however, interactions with a strongly dissipative system have not been included.

\section{Results and Discussion}

\subsection{Entanglement Formation in Weak and Strong Coupling Regimes}

\begin{figure}[t]
\begin{center}
\includegraphics[width=100.mm]{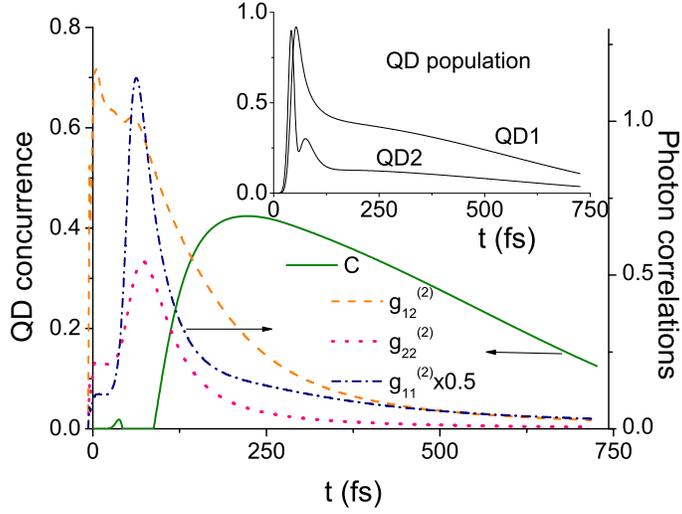}\vspace{-1.0cm}
\end{center} 
\caption{(Color online)  The dynamics of the system for weak QD-cavity photon coupling, $\xi =0.268$: 
QD concurrence $C(t)$ and the same-time cavity photon pair correlation function $g_{ij}^{(2)}(t)$ (main plot), 
and the QD population (inset). 
The system is optimally assembled with the QD-plasmon coupling constants ratio of $1/\sqrt{3}$ that maximizes the QD entanglement; 
$\hbar g_s^i=30$ and 17.3 meV for $i=1$ and 2, respectively;  
$\hbar g=1$ meV; $\hbar \gamma_s = 150$ meV;
the QD decay and dephasing rates are 0.05 $\mu$eV and 8.6 $\mu$eV;  the respective photon
decay and dephasing rates are 0.1 meV and 8.6 $\mu$eV; the transition dipole moments for the surface plasmons and QDs are $d_s = 4000$~D and $d_i=13$~D; the energy level spacing of the QD and cavity photon systems is $\hbar \omega = 2.05$~eV
(Refs.\ \cite{Hennessy:07,Otten:15,Otten:16}).
The maximum electric field in the driving pulse is reached at $t=36.3$ fs.}
\label{fig:weakcoupl}
\end{figure}

We first study the quantum dynamics of a pulsed system (\ref{eq:master}) in the weak coupling regime for the
quantum dot/photonic cavities.
Representative results are shown in Fig.~\ref{fig:weakcoupl} for  $\xi =0.268$. 
(In the estimate of the respective dimensionless coupling, 
we use the average $g_s={1 \over 2}\sum_{i=1,2} g_s^i$ as the characteristic QD-plasmon interaction strength.)
It is seen in the inset of Fig.~\ref{fig:weakcoupl} that the initial QD population oscillations damp at $t\sim 100$~fs after 
the system is excited by the laser pulse; 
at later times the QD populations are not equal to each other due to the difference in 
the QD-plasmon coupling strengths. 
As the result of this asymmetry, the QD entanglement is formed at $t_C\approx 87$ fs and reaches the maximum $C \sim 0.42$ at $t\approx 220$ fs, as seen in the main plot of Fig.~\ref{fig:weakcoupl} (see  Appendix B for more details).
It is  evident from Fig.~\ref{fig:weakcoupl} that both the cross- and same-cavity  correlations $g_{ij}^{(2)}$
of the photons decrease starting from $t\approx t_C$; that is, in the same time domain where $C(t)>0$.  
At later times, $t > 500$ fs, the correlations reach $g_{ij}^{(2)}<0.1$,
corresponding to strong antibunching of the cavity photons.
It is worth noting that the time dependence $C(t)$ shown in  Fig.~\ref{fig:weakcoupl} is qualitatively 
similar to that obtained earlier in the simulations in Refs.\  \cite{Otten:15,Otten:16} 
for plasmonically coupled QDs, for which the spontaneous photon emission  was described as 
dephasing in the respective QD Lindblad operator. 
Thus, we infer that at $\xi \ll 1$, the QD dynamics  results in  the significant suppression of  the cavity photon correlations,
whereas the
photon dynamics mainly contributes to small QD dephasing rates.

\begin{figure}[b]
\begin{center}
\includegraphics[width=85.mm]{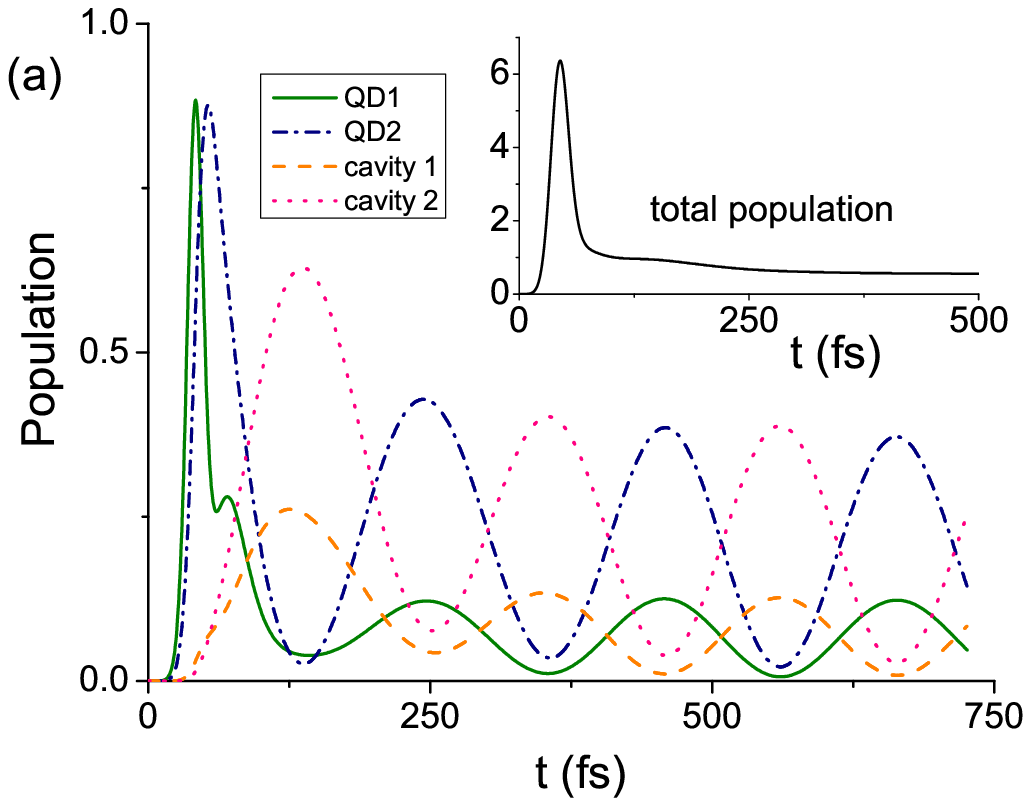}%
\includegraphics[width=85.mm]{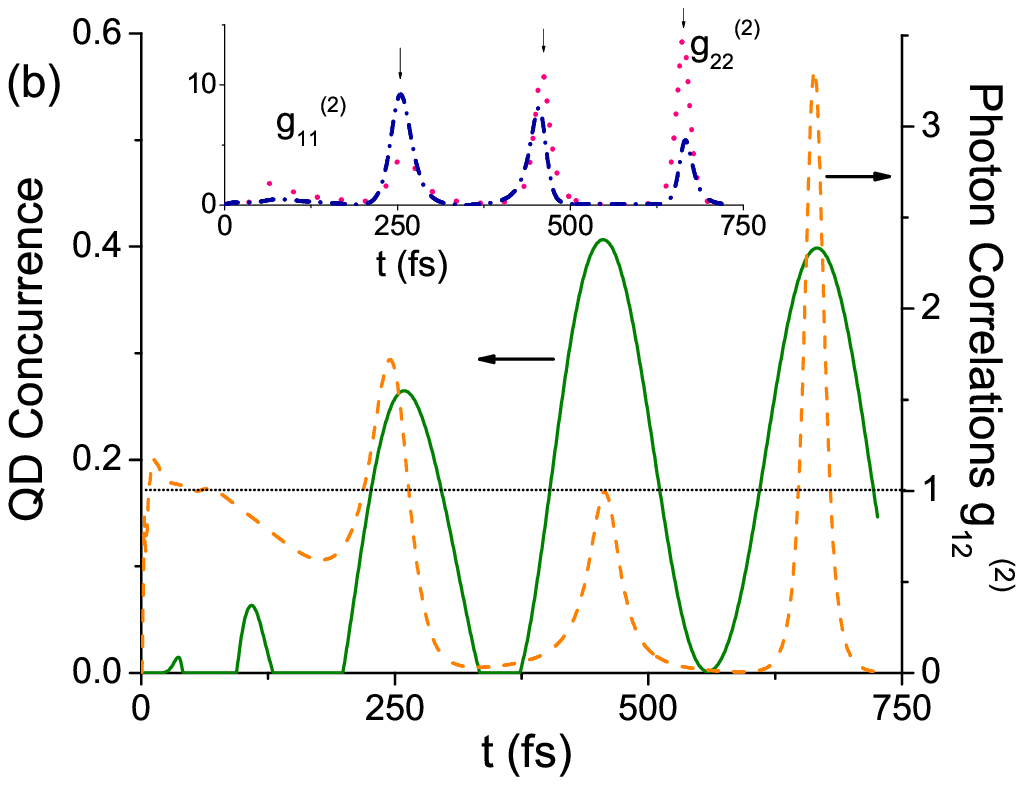}
\end{center} 
\caption{(Color online)  Oscillatory  dynamics of the system in the strong coupling regime $\xi = 2.68$.
(a) Population of QD 1 and 2 and  photon population in cavities 1 and 2. 
Inset shows the total population in the system as a function of time $t$.
(b) QD concurrence and photon correlation functions. Vertical arrows in the inset approximate the moments 
at which the QD concurrence reaches the maxima, as shown in the main plot.
The simulations were done for the same parameters as in Fig.~\ref{fig:weakcoupl}, but 
the QD-cavity photon coupling constant $g$ was increased $10\times$.}
\label{fig:concurrence}
\end{figure}

To study the effect of QD-photon interactions on the quantum dynamics of the system, we increased the coupling  
strength $\xi$. We found that the results for $\xi>1$ significantly differ from those obtained above in the weak coupling regime. 
Figure \ref{fig:concurrence} shows our findings for  large coupling $\xi=2.68$.
It is clearly seen in Fig.\ \ref{fig:concurrence}a in that, 
after being excited by the driving pulse at $t \approx 36.3$~fs, both the QD
and cavity photon populations exhibit  oscillations 
with the period of $t_0 \approx \pi \hbar / g \approx 217$~fs. 
However, the total population in the system does not show any significant oscillations (inset in Fig.\ \ref{fig:concurrence}a).
The main plot in Fig.~\ref{fig:concurrence}b shows formation and subsequent oscillations of the QD   
concurrence $C(t)$ that accompanies the population oscillations in Fig.~\ref{fig:concurrence}a.  
It is seen that starting from $t \sim 200$ fs, $C(t)$ reaches maxima at the same times when
 the QD population builds up. 
 
Figure~\ref{fig:concurrence}b also reveals that, after the initial 150-fs period of relaxation, the cavity photon cross-correlation function 
$g_{12}^{(2)}(t)$ exhibits oscillations that are synchronous with the QD concurrence oscillations.
Specifically, the sharp spikes on the $g_{12}^{(2)}(t)$ curve are positioned 
at the same moments when $C(t)$ reaches its maxima. The 
correlation functions $g_{11}^{(2)}(t)$ and $g_{22}^{(2)}(t)$ for the same cavity photon modes 
follow a similar pattern, as is evident from the  inset in Fig.~\ref{fig:concurrence}b. 
Starting from $t \sim 300$~fs, the cavity photons show strong antibunching with 
$g_{\rm min}^{(2)} < 0.2$ in time intervals between the maxima. 
The relative amplitude $k= (g_{\rm max}^{(2)} - g_{\rm min}^{(2)})/(g_{\rm max}^{(2)} + g_{\rm min}^{(2)})$ 
of the  oscillations reaches $k >0.9$ that makes it accessible for experimental observations. 
(Here, $g_{\rm max(min)}^{(2)}$ are the maximum (minimum) values of $g_{ij}^{(2)}(t)$  for $i,j=1,2$). 

The early-time behavior in Fig.~\ref{fig:concurrence}b does not show the 
interesting correlations between QD concurrence and photon-pair correlation
functions.  This is because of the nature of the experiment we are imagining
that involves an initial pulse 
exciting the QDs followed by 
QD-metal particle interaction and
entanglement via plasmon interactions. These correlations begin only once
a significant concurrence has been established, around 250 fs.

\subsection{Cavity Photon Entanglement and Correlations in the Strong Coupling Regime}

\begin{figure}[b]
\begin{center}
\includegraphics[width=85.mm]{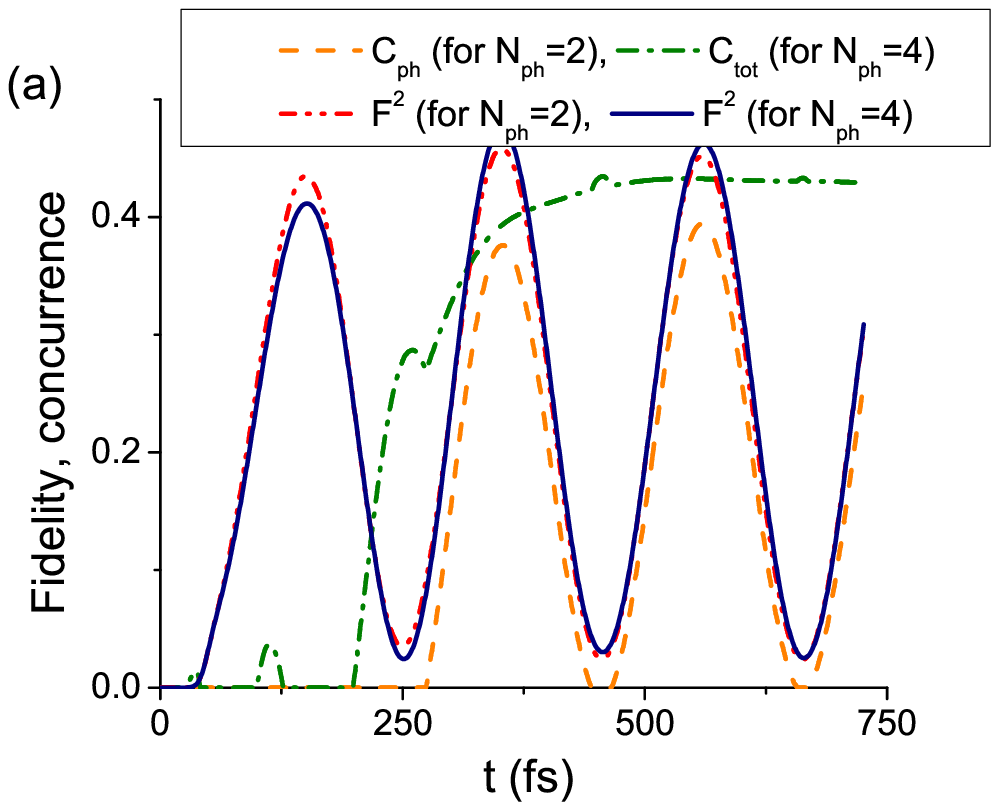}%
\includegraphics[width=85.mm]{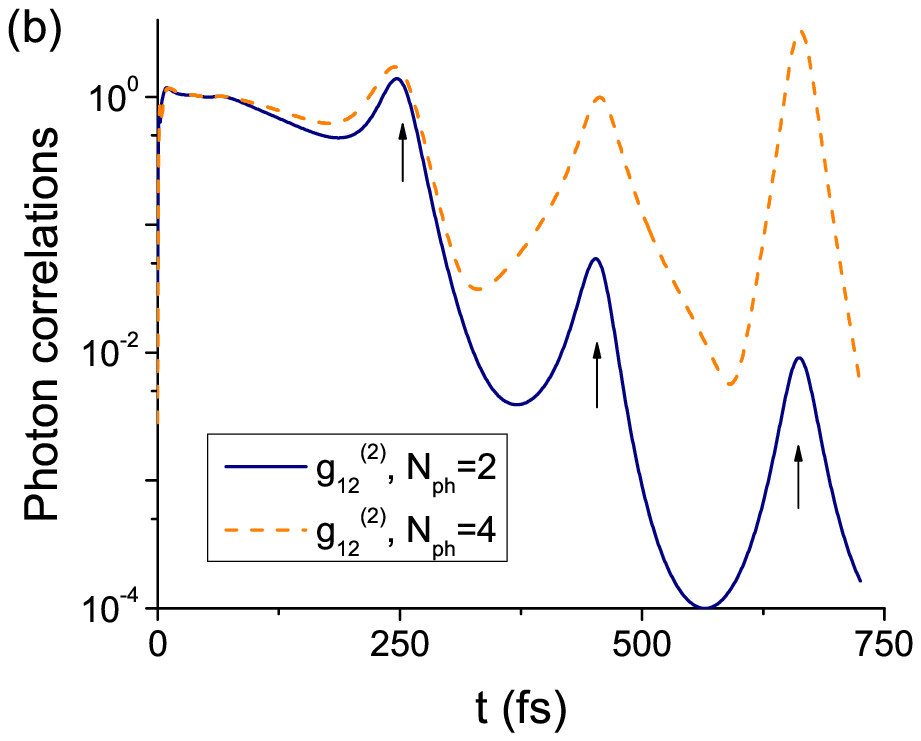}\vspace{-1.0cm}
\end{center} 
\caption{(Color online)  (a) Photon and total entanglement and (b) the photon cross-correlations 
for the model where the number of photon levels is restricted to $N_{ph}=2$. 
The results are obtained for the strong coupling regime with the same parameter set as in Fig.\ \ref{fig:concurrence}.
It is seen in (a) that the squared fidelity $F^2(t)$ of the photon state relative to the maximally entangled Bell state $\Psi^{-}$
follows the photon concurrence. It is  evident from (a) that change in  the photon level number $N_{ph}$ from 2 to 4 does not 
result to significant changes in fidelity $F(t)$ thus, the photons are entangled at $N_{ph}>2$.
As is also seen in (b), the cross-correlation function $g_{12}^{(2)}$ calculated for $N_{ph}=2$ and 4
 shows similar qualitative patterns with sharp peaks positioned at the moments when the QD 
entanglement reaches the maximum values in Fig.\ \ref{fig:concurrence} (arrowed).}
\label{fig:phot}
\end{figure}

To further understand the effect of the oscillatory dynamics on  entanglement, 
we investigated  the correlation properties of the cavity photons. 
For that purpose, we 
restricted the number of energy levels of  the photon cavity modes in the simulations to $N_{ph} = 2$ for both cavities.
This enabled us to 
determine the entanglement  of photons via Wootters' concurrence 
$C_{ph}(t)$ for the reduced photon density matrix along with the QD concurrence $C(t)$.
Our findings are summarized in Fig.\ \ref{fig:phot}. 
It is clearly seen in Fig.~\ref{fig:phot}a that, while the concurrence of the photon modes $C_{ph}(t)$ oscillates with
time similar to that for QDs in Fig.\ \ref{fig:concurrence}a,
the total concurrence $C_{tot}=C + C_{ph}$ shows a smooth time dependence. In other words, 
the entanglement is periodically ``transferred'' between the QD and photon states synchronously with the 
QD and photon population oscillations, with the total entanglement $C_{tot}$ almost conserved  within one oscillation period.
We also numerically calculated the fidelity $F(t)$ of the photon states relative to the  maximally 
entangled Bell state $\Psi^{-}$  that is, to the (antisymmetric) state that mostly contributes 
to the long-time evolution of the system.\cite{Otten:16} It is evident from Fig.~\ref{fig:phot}a
that after $t>250$ fs, $F(t)$ oscillates simultaneously with the photon concurrence $C_{ph}(t)$. 
Thus, $F(t)$ can be viewed as a qualitative characteristic of  the photon entanglement.
We  compared $F(t)$ dependence calculated for $N_{ph}=2$ with that at $N_{ph}=4$
(for which our main results were obtained). 
As is seen in Fig.\ \ref{fig:phot}a, the photon fidelities $F(t)$ for both cases
are close to each other. 
Moreover, it is also evident from Fig.\ \ref{fig:phot}b that, whereas the photon cross-correlations $g_{12}^{(2)}(t)$
calculated for $N_{ph}=2$ and 4 are quantitatively different, they show similar qualitative time dependences with 
sharp peaks positioned at the moments when the QD entanglement reaches its maximum values (arrows).
Therefore, based on the close similarities of the photon fidelities $F(t)$ and the cross-correlation function  $g_{12}^{(2)}(t)$
for $N_{ph}=2$ (for which the concurrence can be explicitly calculated) and $N_{ph}=4$, we infer that 
the cavity photon states emerging in the 
population oscillations are entangled in both cases. We also can say that the $g_{12}^{(2)}(t)$ 
oscillatory time dependence witnesses the underlying QD entanglement.

It should be noted that the concurrence displayed in Fig.~\ref{fig:phot}a
is that associated with the photonic modes and so
will behave in the opposite manner of the QD
concurrence owing to the exchange of entanglement between QD and photonic
modes.

The population and entanglement oscillations observed in the strong coupling regime 
can be attributed to formation of a correlated  QD excitons-cavity photon state, i.e., an exciton polariton. (We will refer to the latter as a polariton.) 
The  polariton state is a quantum superposition of an exciton and a cavity photon, and it 
has been extensively studied in semiconductor quantum well heterostructures embedded  in a 
high-finesse optical microcavity (see Refs.\  \cite{Carusotto:13,Voronova:15} and references therein).
 In our case, however, the excitons -- the ``matter'' part of polaritons -- are localized in QDs.
The polariton  was recently  observed in the strong coupling regime 
in CQED experiments \cite{Hennessy:07} with gallium-arsenide (GaAs) QDs embedded in a photonic crystal nanocavity.
In all these cases, pure exciton and cavity photon states are not the eigenstates of the system. 
If the system is initialized in one of these states, e.g., by the laser excitation, the system exhibits 
Rabi oscillations with the characteristic energy of $\hbar g$. Our studies demonstrate that, if
such a polariton  is formed in  two entangled QDs, the entanglement is also transferred to the 
photon counterpart of the polariton  together with the respective population oscillations.

To obtain deeper insight into  the relation between the photon correlations and QD entanglement, we 
consider  a model where the cavity photon level number is 
again set to $N_{ph}=2$ 
(so that the photon entanglement can be explicitly quantified) and
use notation
$|lm\rangle$ to denote the photon state described by the  
occupation numbers $l(m)$. 
The only long-living state in the system is \cite{Otten:16}
$\Psi^{-} = {1 \over \sqrt{2}}(|01\rangle - |10\rangle )$ and, at the same time,  the cross-correlations between cavity modes 
arise from $\Phi_{11} = |11\rangle$ state. Furthermore, both states relax to the ground state $\Phi_{00} = |00\rangle$ at long times.  Therefore, we only consider the restricted phase space of photons 
that is described by a parametric family 
\begin{equation}
\Phi (x,y) = A( x \Phi_{11} + y  \Phi_{00} +  \Psi^{-}),
\end{equation}
with $x$, $y$ and $A$ being the respective mixing parameters and 
normalization coefficient. 
In this case, immediately after the excitation (that is,  when the occupation of the ground state is small, $y \ll 1$)  one obtains the following relation between the
photon entanglement  and the normalized cross-correlation function,
\begin{equation}
g_{12}^{(2)} = {1 - C_{ph} \over (1 - {1 \over 2} C_{ph})^2 } \quad ({\rm at} \,\, y \ll 1). \label{eq:stepheneq}
\end{equation}
(For the un-normalized  correlations $G_{12}^{(2)} \equiv n_1 n_2 g_{12}^{(2)}$, one has a simple equation $G_{12}^{(2)} = 1 - C_{ph}$.)
For larger times  $t>250$~fs, the photon population is decreasing with time as is seen in Fig.\ \ref{fig:concurrence}.
In the limiting case of small  population of the two-photon state, $x \ll 1$, one has the following for the normalized photon correlation function,
\begin{equation}
g_{12}^{(2)} = {4 x^2 \over C_{ph}} \quad ({\rm at} \,\,  x \ll 1). \label{eq:stepheneq2}
\end{equation}
In other words, in both cases (\ref{eq:stepheneq}) 
and (\ref{eq:stepheneq2}), an increase in  photon entanglement results 
in a simultaneous decrease in the photon cross correlations and vice versa.
Since in the strong coupling regime the photon concurrence is directly transferred from the QD concurrence,
the photon cross-correlations can be seen as a measure of the QD entanglement.
Despite that for $N_{ph}>2$, Eqs.\ (\ref{eq:stepheneq}) 
and (\ref{eq:stepheneq2}) are not  directly applicable, our 
general conclusion remains valid since in the strong coupling regime the total entanglement 
oscillates between the photon and exciton states due to exciton-polariton formation, as it was demonstrated above.
This conclusion confirms the results of our numerical simulations shown in Fig.\ \ref{fig:concurrence}.
It is worth noting that for $N_{ph}=2$, the same-cavity correlations are $G_{11}^{(2)} = G_{22}^{(2)} \equiv 0$.
Thus, the use of two separate cavities is important in this case. In a general case $N_{ph}>2$,  
$G_{11}^{(2)}$ and  $G_{22}^{(2)}$ do not equal zero; however, in the low-excitation mode $n_1 \sim n_2 \sim 1$, one has 
$G_{ii}^{(2)} \ll G_{12}^{(2)}$  due to small occupation of the higher photonic energy levels and thus, an experiment with two individual cavities
is more suitable for experimental observation of the photon correlations compared to that  where two QDs are embedded 
in a single cavity.

\subsection{Entanglement Storage in High-Quality Cavity Modes}

\begin{figure}[b]
\begin{center}
\includegraphics[width=100.mm]{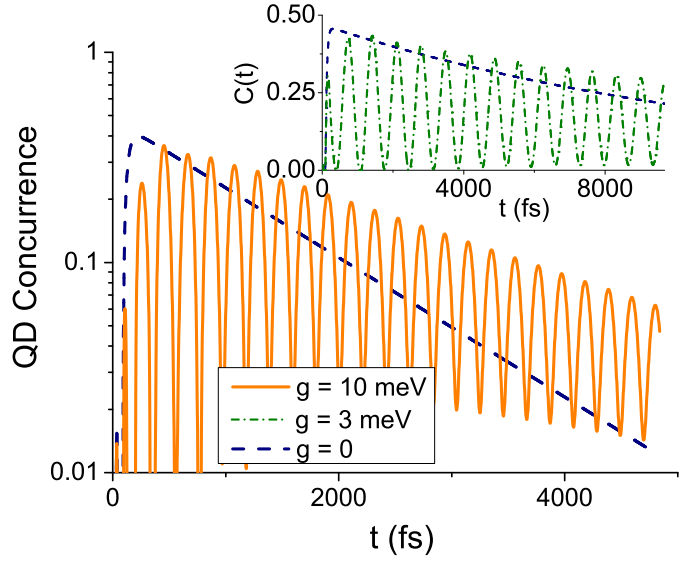}\\
\vspace{-1.0cm}
\end{center} 
\caption{(Color online) Storage of the entanglement of QDs in high-finesse cavities in the strong coupling regime, 
compared to QDs in an open geometry. 
As is seen in the main plot, at $t = 4814$~fs, the concurrence of QDs in the cavities is $\approx 4.58 \times$ greater that that with no cavities.
The QD decay rates are 500~$\mu$eV (main plot) and 50 $\mu$eV (inset); the cavities' quality factor is $Q=10^6$; the photon and QD
dephasing rates are 8.6 $\mu$eV; the cavity photon energy is 
$\hbar \omega = 2.05$~eV; the cavity photon decay rate is $Q^{-1}\hbar \omega =2.05$~$\mu$eV \cite{Knowles:11,Senellart:08,Hennessy:07,Otten:15,Otten:16}.
The strength of the QD-photon interactions is marked in the plot.}
\label{fig:preservation}
\end{figure}

To more fully characterize the effects of coupling with the cavity modes on QD entanglement, 
we compared the time dependencies for the concurrence $C(t)$ of QDs in the cavities in the strong coupling regime with that 
obtained when the cavity modes are absent (an open geometry with $g=0$).
 In these simulations, we take the QD decay rates to be \cite{Knowles:11,Senellart:08}  50~$\mu$eV  and 500~$\mu$eV  and compared the previoulsy obtained results.
Our findings are summarized in Fig.\ \ref{fig:preservation}.
It is seen that, since the characteristic decay rate of photon modes in  high-quality microcavities with $Q\sim 10^6$
are smaller than the QDs non-radiative population relaxation rate,
the QD concurrence time decay rate is weaker in the case when the QD-cavity photon mode coupling is present.
In other words, the entanglement of QDs in the optical cavities  is {\it stored} in the 
high-quality subsystem (photons) for a longer time compared 
to QDs in an open geometry. 
Specifically, 
at $\hbar g = 10$ meV, the concurrence of QDs in the cavities is $\approx 4.58 \times$ greater than 
that in the open geometry at  $t=4814$~fs for the QD relaxation rate of 500 $\mu$eV, 
and is $1.35 \times$ greater at $t=9027$~fs for $\hbar g = 3$ meV and the QD relaxation rate of 50 $\mu$eV,
as is seen in the main plot and inset of Fig.\ \ref{fig:preservation}, respectively.
The effective concurrence decay rate is approximated by the following expression
\begin{equation}
\gamma \approx {\alpha_{QD}} \gamma_{QD} + {\alpha_{C}} \gamma_C, \label{eq:cdec}
\end{equation}
where 
$\alpha_{QD(C)} = { \bar{n}_{QD(C)}  / ( \bar{n}_{QD} + \bar{n}_C )}$ is the fraction of time, 
during which the system occupies the QD (cavity) state, and $\bar{n}_{QD(C)}$ are 
the time-averaged occupation numbers for the QD (cavity). 
Thus, by lowering $\gamma_C$ (increasing the cavity $Q$-factor), one can decrease 
the over-all concurrence decay rate, as follows from Eq.\ (\ref{eq:cdec}). If the QD and cavity modes are in exact resonance, the occupation time average is $\alpha_{QD} \approx \alpha_{C} \approx {1 \over 2}$ and, thus, one has $\gamma \approx {1\over 2} (\gamma_{QD} +  \gamma_C)$. However, under the off-resonant strong coupling conditions, the average occupation time of the cavity mode can
be $\alpha_C > {1 \over 2}$ if the photon-like polariton is excited by the driving pulse.\cite{Carusotto:13}
The latter potentially enables one to further lower the concurrence decay rate.

Finally, we investigated the effect of the pumping pulse duration and of a continuous wave (CW) pump 
on the entanglement formation of asymmetrically coupled QDs
in optical cavities  (see Appendix C). We found that during the period of time when the driving pulse is turned on, the 
QD and photon populations tend to their equilibrium values whereas the QD entanglement does not form in both weak and strong coupling regimes.
However, the QD concurrence $C>0$ formed after $\sim 100$ fs after the driving pulse is turned off.
Therefore, unlike a CW driven system, the free dynamics of the initially excited system could be used to generate and optically detect 
robust QD entanglement.
We also found that setting of the QD-plasmon interaction strength to the same value for both QDs (i.e., where no QD entanglement was formed) 
resulted in $g_{12}^{(2)}(t)  \rightarrow 1$ with no photon antibunching observed, as is detailed in Appendix C.

\section{Conclusion}

In this work, we have shown how to identify entanglement of coupled QD qubits via 
cavity photon correlation measurements. 
Specifically, 
our results could  contribute to quantum simulations that utilize exciton-polariton entangled states \cite{Angelakis:17}.
The obtained results may help one to determine, through  optical experiments, 
the exciton and/or photon entangled states \cite{Benson:00} revealed in recent 
experiments with quasi-two-dimensional core-shell nanoplatelets \cite{Ma:17}.
The conservation of total entanglement we have seen in the strong coupling limit of the cavity/QD qubit interactions  also suggests a novel means of preserving entanglement.

By considering  cavity-photon and QD exciton dynamics coupled with surface plasmons optically excited by 
a femtosecond laser pulse, we showed that the character of the QD entanglement formation is 
different in strong and weak coupling regimes between the photons and QDs -- 
oscillatory  vs slowly decaying entanglement.
In both regimes, the same-time pair correlation function $g_{ij}^{(2)}(t)$ of the cavity photons is sensitive to the QD concurrence formation.
In particular, in the strong photon-QD coupling  regime, $g_{ij}^{(2)}(t)$ peak formation 
-- bunching -- correlates with the QD entanglement formation.
This can be understood as the effect of the entanglement oscillations between the QD  and the cavity photons
due to the exciton polariton  formation.
In the time intervals between the peaks, the photons emitted by the entangled QDs strongly antibunch, $g_{ij}^{(2)}(t) < 0.2$.  This behavior contrasts with $g_{ij}^{(2)}(t) \approx 1$ for unentangled QDs, enabling direct optical detection of QD qubit entanglement. 

The correlations exhibited in our proposed quantum dot/photonic cavity system
can seem surprising.  For example, often non-classical (e.g., entangled) 
states are associated with anti-bunching or small values of pair correlation
function $g^{(2)}$
and this has been shown for plasmon-QD systems
\cite{Dumitrescu:17}.  (See, however, different 
behavior for Gaussian squeezed states \cite{Stobinska:2005}.)
However, keeping in mind we are engineering strong coupling between the
photonic modes and the QDs, there is an exchange of entanglement
between these subsystems. Thus, when the QDs are non-classical
the photonic modes are not, and vice versa, leading to the photon pair correlation
function paralleling the concurrence behavior of the QDs.

To experimentally achieve the solid-state photonic qubit system (Fig.\ 1) proposed one could use, as quantum dots, the cadmium selenide (CdSe) nanoplatelets of Ref.\ \cite{Ma:17}.  While metal nanoparticles represent one avenue for the mode coupling the two QDs, a silicon microdisk resonator supporting a weakly dissipative mode might be more easily used.  The two photonic cavity modes could be realized with high quality photonic crystals composed of silicon nitride, for example.  Finally, one can envision optical fibers connected to the two photonic crystals that would lead to photon coincidence detectors for the time-resolved pair correlation function measurements \cite{Yamamoto:99,Scully:97}.

\section*{ACKNOWLEDGEMENTS}

We thank Xuedan Ma for helpful discussions.  G.V.K. gratefully
acknowledges support from the U.S. Department of Energy, Office of Science, Office of Workforce Development
for Teachers and Scientists (WDTS) under the Visiting Faculty Program (VFP). This work was performed, in part, at the Center for Nanoscale Materials, a U.S. Department of Energy Office of Science User Facility, and supported by the U.S. Department of Energy, Office of Science, under Contract No.\ DE-AC02-06CH11357. 

\appendix

\section{Simulation parameters}

 The pulse intensity is characterized by the
fluence  \cite{Shan:13,Otten:15,Otten:16}
\begin{equation} 
F= \int_{-\infty}^{\infty}  dt  \sqrt{\epsilon_{\rm med}} c \epsilon_0 E^2(t),
\end{equation}
where  $\epsilon_{\rm med}=2.25$ is the relative dielectric constant of the
surrounding polymer matrix, $c$ is the speed of light in vacuum, and $\epsilon_0$ is the vacuum permittivity. In the simulations, we restricted the number of plasmon (cavity photon) energy levels of the underlying physical system Hamiltonian $\hat{H}_0$  to $N_{ph(pl)}$ and then numerically
integrate Eq. (1) in the main text. 

The effect of the electromagnetic interactions of the QD excitons with the damped surface plasmon 
system in a neighboring metal particle or nanostructure is two-fold. First, the asymmetry in the QD-plasmon coupling, 
$\Delta g_s \neq 0$  
induces a spontaneously formed entanglement of the QD excitonic states \cite{Hou:14,Otten:16} 
with the maximum entanglement achieved at $g_s^1/g_s^2 \approx 1/\sqrt{3}$ \cite{Otten:15}.
Here, the asymmetry in the QD-plasmon coupling strength is defined as
\begin{equation}
\Delta g_s \equiv g_{s}^1-g_{s}^2,   \label{eq:deltag}
\end{equation}
 the upper index $i=1,2$ in $g_s^i$ marks the quantum dots, as defined in the main text.
 
Second, due to the Purcell effect, the interactions modify the exciton 
non-radiative decay rate, compared to that in an isolated dot, to $\gamma_{QD} = 4 (g_s)^{2}/\gamma_s$ \cite{Shan:13}
(with $g_s$ to be the averaged QD-plasmon interaction strength, as defined in the main text). 
The latter results in the modified effective coupling constant ($\gamma_{QD} \gg \gamma_C$)
\begin{equation}
	\xi = {g \gamma_{s} \over  g_s^{2}} \label{eq:xi}
\end{equation}
for the symmetric, superradiant exciton states in the coupled QDs. 
However, the antisymmetric, subradiant collective exciton states are only weakly coupled to the plasmonic system \cite{Otten:15} thus,
the non-radiative decay $\gamma_{QD}$ dominates in this case. 
In the pulsed pumping, 
the symmetric and antisymmetric states are initially excited, but the symmetric state rapidly 
decays due to coupling with surface plasmons thus, controlling the fast population damping mechanism in the system.
Thus, we characterize our system via the effective coupling constant $\xi$, as defined in Eq.\ (\ref{eq:xi}) above.

\section{Symmetrically coupled quantum dots}

\begin{figure}[h]
\begin{center}
\includegraphics[width=100.mm]{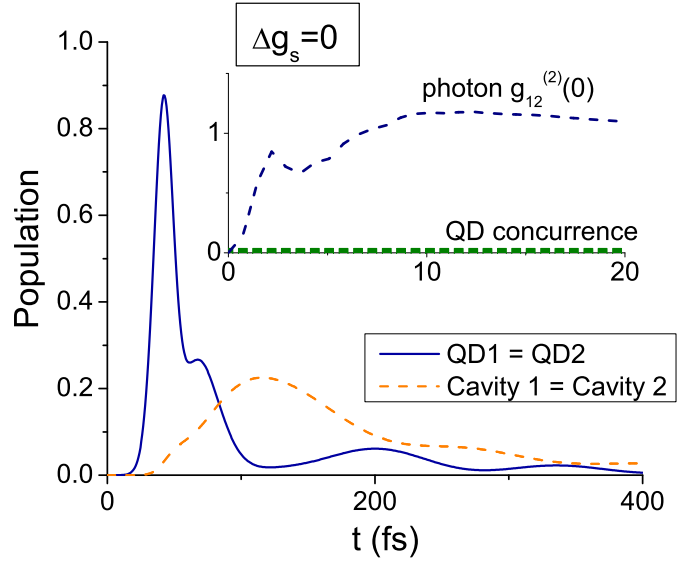}\vspace{-1.0cm}
\end{center} 
\caption{(Color online) (a) Population (main plot), QD concurrence and cavity photon correlations (inset) for equal plasmon coupling, $\Delta g^s = 0$.
It is seen in the inset that the QD concurrence is $C(t) \ll 1$ and, 
at the same time, the pair cross-correlation function for the cavity photons tends to $g_{12}^{(2)}(t) \approx 1$ for $t>10$ fs. 
The QD-plasmon interaction strength is the same for both dots, $\hbar g_s^1= \hbar g_s^2=30$ meV;  
$\hbar g=1$ meV; $\hbar \gamma_s = 150$ meV;
the QD decay and dephasing rates are 0.05 $\mu$eV and 8.6 $\mu$eV;  the respective photon
decay and dephasing rates are 0.1 meV and 8.6 $\mu$eV; 
the transition dipole moments for the surface plasmons and QDs are $d_s = 4000$~D and $d_i=13$~D; 
the energy level spacing of the QD and cavity photon systems is $\hbar \omega = 2.05$~eV.
}
\label{fig3:equalgs}
\end{figure}

To  demonstrate that the formation of the QD entanglement is the key factor influencing the
cavity photon correlation pattern described above,
we studied the dynamics of the system with symmetrically coupled QDs,  that is with $\Delta g_s = 0$.  It is known that 
the concurrence formed in plasmonically coupled QDs is negligible  when the coupling values are equivalent \cite{Otten:15,Otten:16}.
Our results obtained for $\xi_{\rm eff}=2.68$ are shown in Fig.~\ref{fig3:equalgs}.
It is seen that in this case 
the QD concurrence is $C(t) < 0.06$ in accordance with 
the existing theory and simulations in Ref.~\cite{Otten:15,Otten:16}. 
It is also evident from  Fig.~\ref{fig3:equalgs} that the cavity photon cross-correlation function $g_{12}^{(2)}(t)$
rapidly approaches unity.

\section{Continuous wave pumping}
To study  the effect of the CW pumping on the system dynamics, we simulate the quantum dynamics of the system 
under very long laser pulse durations up to 700 fs.
In these simulations, the semiclassical electric field of the laser pump  is taken to be $E(t)=E_0(t) \cos (\omega t)$ with the envelope function
\begin{equation}
 E_0(t) = E_{\rm max} {( \tanh[ (t_c-t_0)/ \delta] +1 )^{-1}  + (\tanh[ (t_1-t_c)/ \delta] +1 )^{-1} \over ( \tanh[ (t-t_0)/ \delta] +1 )^{-1}  + (\tanh[ (t_1-t)/ \delta] +1 )^{-1} } . \label{eq:pulse}
\end{equation}
where $E_{\rm max}$ is the maximum value of the electric field  in the pulse, and $t_c = {1\over 2} (t_1-t_0)$ marks the middle 
of the time domain where the pulse is applied. For the pulse duration $\Delta t \equiv t_1-t_0 = 20$ fs and width $\delta=10$~fs, the pulse (\ref{eq:pulse}) approximates the Gaussian pulse
that we used in the main text. 

\begin{figure}[h]
\begin{center}
\includegraphics[width=80.mm]{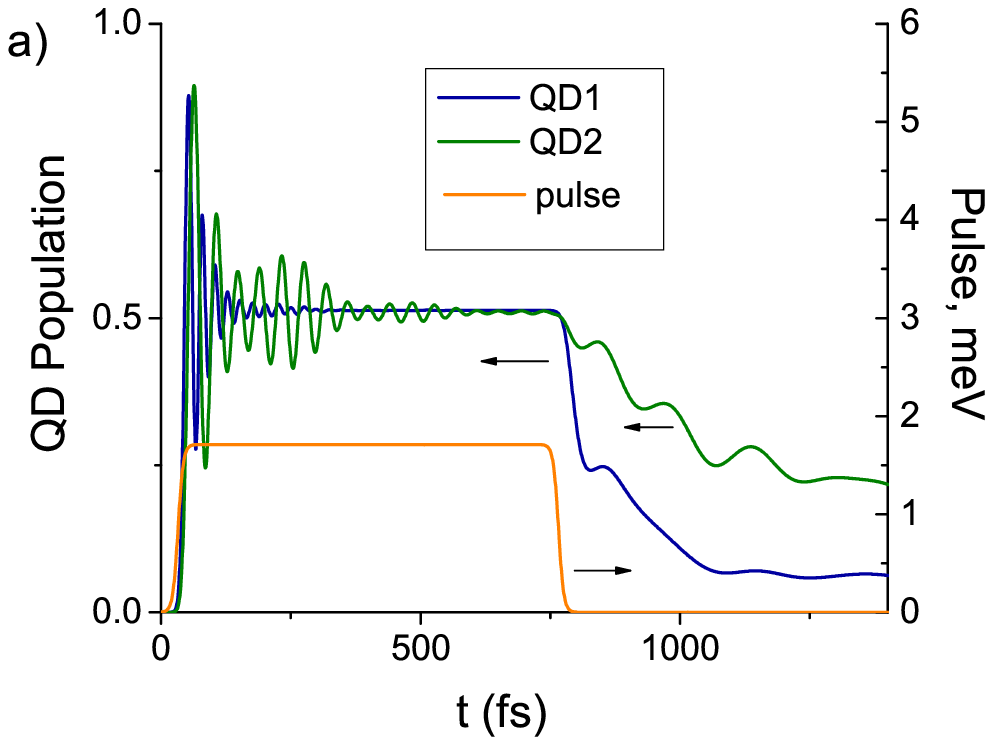}
\includegraphics[width=80.mm]{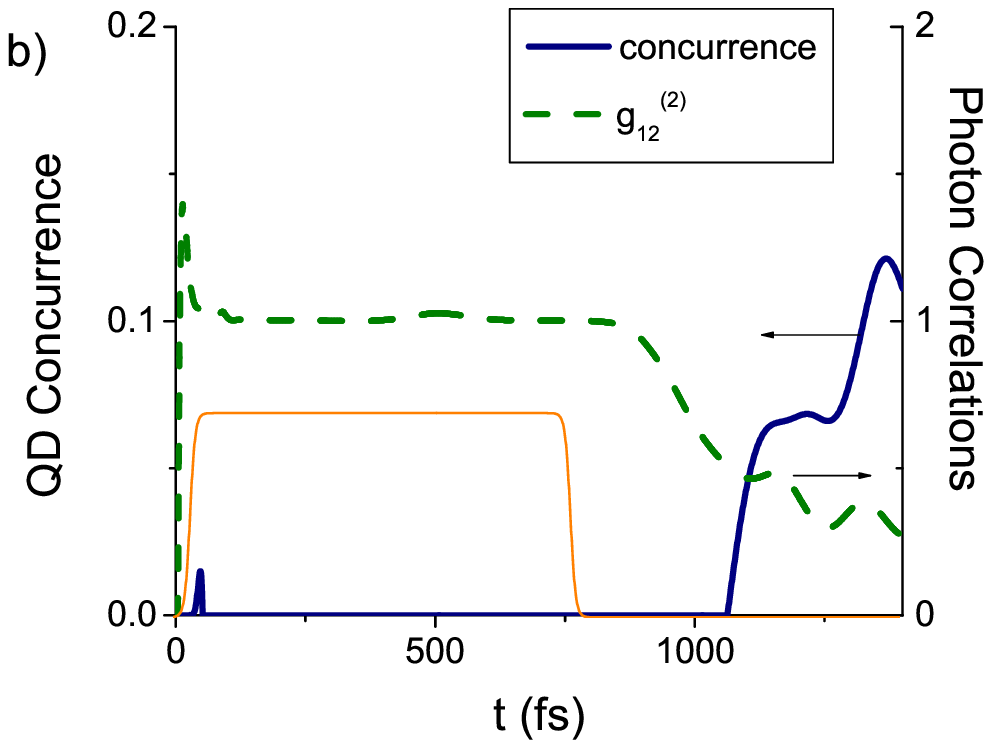}
\end{center} 
\caption{(Color online) Quantum dynamics of plasmonically coupled QDs and cavity photons under the action of a long laser pump.
The time duration of the laser pulse is $\Delta t = 720$ fs, the characteristic pulse formation and decay time is 
$\delta = 10$ fs. The maximum electric field in the pulse is $E_{\rm max} = 2.5\times 10^6$ V/m. (a) Population of the 1st (QD1) 
and 2nd (QD2) quantum dots as functions of time $t$ (left scale). The orange curve shows the  shape of the envelope of the laser pulse
$E_0(t) d_i$ where $d_i = 13$~D is the  transition dipole moment of QD and $E_0(t)$ is given in Eq.\ (\ref{eq:pulse}) (right scale). It is seen that after the transient oscillations are damped, the QD populations tend to a steady-state value of 0.5 when the laser pump is turned on  and then
decrease with time after the pump is turned off.
(b) The QD concurrence (left scale) and the cavity photon cross-correlation function $g_{12}^{(2)}$ (right scale). 
The orange curve shows the laser pulse envelope curve (in arb. units).
It is seen that the QD concurrence is $C(t)=0$ when the pump pulse is on and then 
begins to increase $\approx 180$ fs after the driving pulse was switched off. The photons are emitted with  $g_{12}^{(2)} \approx 1$ 
when the pulse is switched on.
The photons antibunch,  $g_{12}^{(2)} < 1$, when the QD entanglement is formed after the driving pulse is turned off. 
In the simulations, the QD-plasmon coupling strength  are 
30 meV and 17.3 meV for QD1 and QD2, respectively; the QD-photon coupling strength is 10 meV; the plasmon decay rate is 150 meV;
the QD decay and dephasing rates are 0.05 $\mu$eV and 8.6 $\mu$eV, respectively;  the respective photon
decay and dephasing rates are 0.1 meV and 8.6 $\mu$eV; 
the transition dipole moments for the surface plasmons and QDs are $d_s = 4000$~D and $d_i=13$~D; 
the energy level spacing of the QD and cavity photon systems is $\hbar \omega = 2.05$~eV.
}
\label{fig:cw}
\end{figure}

Here we vary the pulse duration $\Delta t$ from 20 fs to 720 fs.
In all simulations, we observe that within the time domain where the pulse is applied, $t_0 - \delta < t < t_1+\delta$, the QD concurrence is 
equal to zero, and the cavity photon correlation function was $g_{12}^{(2)} \approx 1$.
A typical output of the simulations is shown in Fig.~\ref{fig:cw}.
We also find that approximately $100-300$ fs after the pulse is switched off, the entanglement of the QD states occurs (Fig.\ \ref{fig:cw}b). 
At the same time, the photon correlation function is decreased, $g_{12}^{(2)} < 1$, which further confirms our conclusion about 
the photon antibunching in states with 
entangled QDs. 

We did not observe formation of the QD entanglement within 
the time domain $t_0 - \delta < t < t_1+\delta$ when the pulse is turned on.
Thus, we infer that the presence of the external CW laser pumping destroys the entanglement in the system and, at the same time,  
results in virtually coherent emission of cavity photons by the QDs with $g_{12}^{(2)} \approx 1$. 
This conclusion is in qualitative agreement with the results of existing 
numerical and analytic analyses of the plasmonically coupled QD dynamics in Ref.\ \cite{Otten:16}.


%

\clearpage

\end{document}